\documentclass[referee]{aa}  

\usepackage{graphicx}

\usepackage{txfonts}
\usepackage{xcolor}
\usepackage[colorlinks=true, citecolor=blue]{hyperref}

\begin{document}

   \title{Rubidium abundances in solar metallicity stars}
   
    \titlerunning{Rubidium abundances in solar metallicity stars}
   
   \authorrunning{C. Abia et al.}

   \author{C. Abia\inst{1}, P. de Laverny\inst{2}, S. Korotin\inst{3}, A. Asensio Ramos\inst{4,5}, A. Recio-Blanco\inst{2}, N. Prantzos\inst{6}}
    
     \institute{Dpto. F\'\i sica Te\'orica y del Cosmos. Universidad de Granada, E18071 Granada, Spain\\
              \email{cabia@ugr.es}
              \and 
              Universit\'e C\^ote d’Azur, Observatoire de la C\^ote d’Azur, CNRS, Laboratoire Lagrange, France
              \and
              Crimean Astrophysical Observatory, Nauchny 298409, Republic of Crimea 
              \and
              Instituto de Astrof\'isica de Canarias (IAC), Avda V\'ia L\'actea S/N,
                38200 La Laguna, Tenerife, Spain
         \and
             Departamento de Astrof\'isica, Universidad de La Laguna, 38205 La Laguna, Tenerife, Spain
          \and 
          Institut d’Astrophysique de Paris, UMR7095 CNRS \& Sorbonne Universit\'e, 98bis Bd. Arago, F-75104 Paris, France
             }

   \date{Received; accepted }

 \abstract
   {Rubidium is one of the few elements  produced by the neutron capture s- and r-processes almost in equally proportions. Recently, a Rb deficiency ([Rb/Fe]~$<0.0$) by a factor of about two with respect to the Sun has been found in M dwarfs of near solar metallicity. This contrasts with the close to solar [Sr,Zr/Fe] ratios derived in the same stars. This deficiency is difficult to understand from both the observational and nucleosynthesis point of views.}
   {To test the reliability of this Rb deficiency, we study the Rb and Zr abundances in a sample of KM-type giant stars in a similar metallicity range extracted from the AMBRE Project.}
   {We use high resolution and high signal-to-noise spectra to derive Rb and Zr abundances in a sample of 54 bright giant stars with metallicity in the range $-0.6\lesssim$[Fe/H]$\lesssim +0.4$~dex by spectral synthesis in both local and non-local thermodynamic equilibrium (LTE and NLTE, respectively). The impact of the Zeeman broadening in the profile of the \ion{Rb}{i} at $\lambda 7800$ {\AA} line is also studied.}
   {The LTE analysis also results in a Rb deficiency in giant stars although considerably lower than that obtained in M dwarfs. However, when NLTE corrections are done the [Rb/Fe] ratios are very close to solar (average $-0.01\pm 0.09$ dex) in the full metallicity range studied. This contrasts with the figure found in M dwarfs. The [Zr/Fe] ratios derived are in excellent agreement with those obtained in previous studies in FGK dwarf stars with a similar metallicity. We investigate the effect of gravitational settling and magnetic activity as possible causes of the Rb deficiency found in M dwarfs. While, the former phenomenon has a negligible impact on the surface Rb abundance, the existence of an average magnetic field with  intensity typical of that observed in M dwarfs may result in systematic Rb abundance underestimations if the Zeeman broadening is not considered in the spectral synthesis. This can explain the Rb deficiency  in M dwarfs, but not completely. On the other hand, the new [Rb/Fe]  and [Rb/Zr] vs. [Fe/H] relationships can be explained when the Rb production by rotating massive stars and low-and-intermediate mass  stars (these later also producing Zr) are considered,  without the need of any deviation from the standard s-process nucleosynthesis in AGB stars as suggested previously.}
{}

\maketitle
%

\section{Introduction}
The chemical evolution of the galaxies can be traced through abundance determinations in long-lived FGK dwarfs belonging to the different stellar populations. Theoretically, these stars preserve unaltered in their atmospheres the chemical composition of the original cloud from which they formed. Nowadays a number of large  spectroscopic surveys in the Milky Way devoted mainly to the study of these stars like Gaia-ESO \citep{gil12,jac15}, GALAH \citep{sil15,bud18}, AMBRE \citep{AMBRE13}, APOGEE \citep{ahu20} and others, are providing a huge quantity of spectroscopy data. Together with  the accurate distances and kinematic information determined by the Gaia mission \citep{bro18} followed by accurate stellar age estimations from large asteroseismic surveys         \citep[e.g.][]{mig20},  these studies  are revolutionising the current understanding of the Milky Way history. However, abundance analyses of FGK dwarfs   do  not always allow an easy determination of the abundances of some elements. This is the case for several heavy elements (A $>70$) produced mainly by neutron capture reactions through the s- and r-processes in different astrophysical scenarios \citep[see e.g.][]{bus99,kap11,thi17,cow19}. The universal low abundances of these elements and the physical parameters of the atmospheres of FGK dwarfs usually make their available spectroscopic lines very weak, which in addition are often heavily blended, in particular in stars with near solar metallicity or higher 
\citep[see e.g.][]{jof19}. This problem is worsened when medium resolution spectra are used ($R\lesssim 20000$) as is the case in some of the surveys mentioned above.

Rubidium is one of these elements. Analysis of the rubidium abundance in the Solar System shows that the neutron capture s- and r-processes are about equally responsible for the
synthesis of this element \citep[e.g.][]{sne08,pra20}. Astronomical detection of Rb mainly relies on two resonance \ion{Rb}{I} lines at $\lambda\lambda$ 7800 and 7947~\AA.  In solar-like stars these Rb lines are weak and heavily blended, which makes  difficult an accurate determination of the Rb abundance. In fact, a controversy on the photospheric Solar Rb abundance existed until recently  \citep{gol60,lod09,asp09,gre15}. To date, only two Rb abundance studies FGK-type stars exist, namely those of \citet{gra94} and \citet{tom99}. These authors derived Rb abundances in a sample  of  metal-poor disc and halo stars and,  considering the results in both studies together\footnote{Note that
\citet{gra94} derived  upper limits for the Rb abundance in some of the
stars in their sample., one may concluded} that the [Rb/Fe] versus [Fe/H] relationship behaves at low metallicity ([Fe/H]~$<-1$\footnote{Here we follow the standard abundance notation, [X/H] $= \log{({\rm X/H})_\star} - \log{({\rm X/H})_\odot}$, where X/H is the abundance by number of the element X, and $\log \epsilon(X) \equiv \log{({\rm X/H})} + 12$.}) similarly as [Eu/Fe] does, that is, showing an approximately constant [Rb/Fe] ratio as a typical r-process element. However, the behaviour of [Rb/Fe] at higher metallicities ([Fe/H]~$>-0.5$) was poorly studied because of  the issues commented above. 

An alternative to FGK dwarfs for Galactic chemical tagging are M dwarfs. Due to their ubiquity and very long main-sequence lifetimes, abundance determinations in M dwarfs are a powerful and  complementary  tool to study the formation and chemical enrichment of the Galaxy. Their potential is only beginning to be explored \citep[see e.g.][]{sou20,bir20}. Because of their  low effective temperature (T$_{\rm{eff}}\lesssim 3800$ K), the spectra of M dwarfs usually show strong \ion{Rb}{I} lines, which can be easily identified out of a forest of molecular absorptions (mainly TiO) even in metal-rich stars. Very recently \citet{abi20} (hereafter Paper I) derived Rb abundances for the first time in a sample of nearby M dwarfs in the metallicity range $-0.5\lesssim$[Fe/H]$\lesssim +0.5$ using very high resolution visual and near infrared CARMENES spectra \citep{qui18,Rei18}. In this study Sr, and Zr - two neighbour elements with mainly a s-process origin - were also derived as cross check elements to evaluate the reliability of abundance determinations in M dwarfs. In fact, while the [Sr,Zr/Fe] ratios derived by these authors were in excellent agreement with those observed in FGK dwarfs of similar metallicity \citep[e.g.][]{bat16,del17}, they found [Rb/Fe] ratios systematically lower than solar (i.e. [Rb/Fe] $<0.0$) by a factor two on average, and also a possible trend of increasing [Rb/Fe] ratios for [Fe/H] $>0.0$. These are surprising results never found for any other heavy element at similar metallicities. These authors discussed several possible explanations for these findings in terms of deviations from LTE \citep{Korotin20}, an anomaly of the Rb abundance in the Solar System \citep[e.g.][]{wal09,rit18}, the stellar activity in M dwarfs, and/or a deviation from the standard s-process nucleosynthesis scenario for Rb in AGB stars \citep{cristallo2009,karakas2010,cristallo2018}, but no plausible solution 
was found\footnote{The [Rb/Fe] ratios derived by  the \citet{tom99} in K dwarf stars, which have slightly larger masses than M dwarfs, apparently do not show any systematic difference when compared with the ratios derived in G dwarfs and giants in their stellar sample.}.  In Paper I, it  was finally suggested that  additional  Rb  abundance  measurements  in FGK dwarfs and giants of near solar metallicity, as well as a more detailed evaluation of the impact of stellar activity on abundance determinations in M dwarfs, were urgently needed to confirm or disprove these findings. 

In the present study, we derive Rb abundances from high-resolution spectra in a sample of nearby and bright K and M stars located on the subgiant and giant branches with metallicities close to solar. We also determine their Zr abundance - an element with predominantly main s-process origin - as a cross check to the analysis. For that purpose, we use high signal-to-noise template spectra of 54 giants  provided by the AMBRE Project. Our aim is to study the reliability of the low [Rb/Fe] ratios found in the previous study on nearby M dwarfs of similar metallicity for a better understanding of the evolution of the Rb abundance in the Galaxy, and to put constraints on the role played by the s- and r-processes in the galactic Rb budget. 

The structure of this paper is as follows: the observational material and analysis is presented in Sec. 2, where the data acquisition
is briefly described; we also discuss the atmospheric parameters used in this study, the line lists, and the derivation of the abundances from the spectra, together with an evaluation of the observational and analysis uncertainties. 
In Sec. 3 the main results are presented. The results are then  compared with recent nucleosynthesis models through a state-of-the-art  galactic chemical evolution model for the solar neighbourhood. We also briefly discuss gravitational settling and magnetic activity as  possible explanations of the Rb deficiency found in M dwarfs. Section 4 summarises the main conclusions of this study.
 
\section{Observations and analysis}

\subsection{The stellar sample}
We have looked for ESO-archived UVES spectra collected with the RED860 setup (appropriate
for the observation of the \ion{Rb}{i} lines around $\lambda\lambda$ 7800 {\AA} and 7947~\AA),
and already analysed within the AMBRE Project \citep{AMBRE13}. These spectra have been automatically parametrised within AMBRE using the projection method MATISSE \citep{rec06}, trained on a specific grid of high-resolution synthetic spectra \citep{AMBRE12}. 
The parametrisation of the AMBRE-UVES sample is detailed in \cite{wor16}. It provides, among others,
the stellar radial velocity, the signal-to-noise ratio (SNR), 
the main atmospheric parameters:
effective temperature $T_{\rm eff}$, the surface gravity (log $g$) and the
mean metallicity [M/H], adopted hereafter as an estimate of [Fe/H] and, the enhancement in $\alpha$-elements with respect to iron ([$\alpha$/Fe]).
A quality-flag of the stellar parametrisation, based on the computation of a $\chi ^2$ between the observed and reconstructed spectra at the derived stellar parameters, has also been estimated.

Within these parametrised AMBRE-UVES spectra, we selected only cool ($T_{\rm eff}< 4500$ K) stars belonging to the red giant and sub-giant branches (log g $< 3.0$) with SNR $>100$, to make sure that the Rb lines are clearly detectable. 
Only spectra with a good parametrisation flag ($\le$ 1) were
also considered.
With these criteria, an initial sample of 80 objects were  selected. However, we filtered again the sample excluding those objects with peculiar spectral types (e.g. R-stars, Ap-stars, symbiotics..), those belonging to stellar clusters, and those that might be placed in the asymptotic giant branch (AGB) phase\footnote{It is indeed very well known that AGB stars may show Rb enhancements produced by the in-situ operation  of the main s-process \citep{abi01,gar06}.}, all this according to the SIMBAD database. The final selected spectra set consisted in 54 giant stars of spectral types K and M with SNR~$>150$, most of them  widely studied in the literature for other purposes (see Table 1).

\subsection{Determination of Rb and Zr abundances}
In a first step, we adopted the stellar parameters derived by \citet{wor16}. We refer to this study for a detailed description of the method. We also adopted initially for all the stars a microtubulence parameter $\xi=1.7$ km s$^{-1}$, a typical value for giant stars. We build a model atmosphere  for each star interpolating within the grid of MARCS model atmospheres by \citet{gus08} for the corresponding stellar parameters. Then, a synthetic spectrum calculated in LTE with the Turbospectrum v19.1 code \citep{ple12} was compared with the observed spectrum of each star in specific spectral ranges to check the validity of the stellar parameters. These ranges were about fifty angstroms  centred at $\lambda\lambda\sim 6700$, and 7100 {\AA} and the full range $\lambda\lambda 7750-8100$ {\AA}. Obviously, they include the $\lambda 7800$, and $\lambda 7947$ {\AA} \ion{Rb}{i} lines of interest here, but also several metallic and molecular lines (TiO and CN). These metallic lines served as a check of the metallicity initially adopted in the atmosphere model, while molecular lines were used to estimate the C/O ratio, which determines critically the shape of the spectrum for stars cooler than $T_{\rm eff}\lesssim 4000$ K.

Synthetic spectra were convolved with a Gaussian function having a FWMH in the range 6-9 km s$^{-1}$, to account for the instrumental profile and macroturbulence. 
The atomic line list was taken from the VALD3 database adopting the corrections  performed  within  the Gaia-ESO  survey \citep{hei15b,hei20} in the wavelength ranges
studied. Additional corrections to the log~$gf$ values of specific atomic lines were made from comparison of a synthetic spectrum with the observed spectrum of Arcturus \citep{Hinkle95}. Molecular line lists were provided by B. Plez\footnote{These molecular line lists are publicly available at https://nextcloud.lupm.in2p3.fr/s/r8pXijD39YLzw5T, where detailed bibliographic sources can be also found.} and they include several C- and O-bearing molecules (CO, CH, CN, C$_2$, HCN, TiO, VO, H$_2$O) and a few metallic hydrides (FeH, MgH, CaH). As mentioned above, for the stars with $T_{\rm eff}\lesssim 4000$~K the C/O ratio plays an important role in the shape of the spectrum and, in fact, determines the intensity of a veil of TiO lines present in the spectral regions of the \ion{Rb}{i} lines, which may depress significantly the spectral pseudo-continuum there. To estimate this ratio we proceed as follow: 

i) We scaled the CNO abundances to the  initial metallicity adopted since [C,N,O/Fe]$\approx 0.0$ dex is fulfilled for stars with near solar metallicity as those studied here. Only for a few stars in the sample with a mild metal-deficiency (see Table 1), we adopted some oxygen enhancement according to the accepted relationship [O/Fe]$\approx -0.36$[Fe/H] dex \citep[see e.g.][]{edv93}. According to \citet{wor16}, the overwhelming majority of the stars studied here show no $\alpha$-enhancement or very small ([$\alpha$/Fe]$\lesssim 0.1$ dex). We checked that an $\alpha$-enhancement within this range of values has no effect on the analysis. 

ii) Then the carbon abundance was estimated using some weak CN lines in the $\lambda 8000$ {\AA} region \citep[see e.g.][]{bro89}. Because these lines are slightly sensitive to the $^{12}$C$/^{13}$C ratio, we adopted a $^{12}$C$/^{13}$C$\sim 20$ ratio (this value is typically observed in giant stars of near solar metallicity  after the first dredge-up, see e.g. \citet{char94}) for all the stars, except for those for which measurements were available in the literature. Ideally, to determine the carbon abundance from CN lines, the N abundance should be known a priori from an
independent spectral analysis (e.g. NH or \ion{N}{I} lines), but unfortunately the available spectral region in our spectra do not contain any such lines, nor there are accurate N abundance determinations available in literature for most of the stars in our sample. Therefore, we adopted the N abundance scaled with the stellar metallicity. Note however, that the CN lines used are not very sensitive to moderate variations of the N abundance, in particular in stars with T$_{\rm eff}> 4200$ K. We checked that changes up to $\pm 0.2$ dex in the N abundance adopted and variations of the $^{12}$C$/^{13}$C value within $30\%$, have a minimal
impact in the determination of Rb and Zr abundances.

iii) With this carbon abundance, the oxygen abundance was estimated from fits to TiO lines mainly in the $\lambda 7100$~\AA \ region.

iv) Once the C/O ratio was estimated, we determined again the carbon abundance from the $\lambda 8000$ {\AA} region and the operation was repeated until convergence was reached. For most of the stars, a few iterations were needed.  We estimate an uncertainty in C/O  from 0.05 to 0.1, depending on the specific stellar parameters: in the cooler stars of the sample the uncertainty is lower since CN and TiO lines become more intense as the effective temperature decreases and are more sensitive to changes in the carbon or the oxygen abundance, respectively. Considering this uncertainty, most of the C/O ratios derived are close to the photospheric solar value, (C/O)$_\odot=0.57\pm 0.04$ \citep{lod19} or slightly lower, this later is that being expected after the operation of the first dredge-up (see Table 1).

For most of the stars we find a nice agreement between observed and synthetic spectra in all the spectral ranges mentioned above. This procedure served also as a test to validate the stellar parameters of the stars according to the estimations within the AMBRE Project. However, for some stars, small discrepancies between observed and theoretical spectra  were detected indicating that some stellar parameters derived in AMBRE may slightly depart from the ones fitting better the shorter wavelength ranges of the present study, in particular the $T_{\rm eff}$. For these stars, we searched in the most recent literature other estimations of the stellar parameters and tested them in the same way as described above until an agreement between the observed and theoretical spectrum was found. In average effective temperatures finally adopted differed from those in AMBRE by $\sim -53\pm 100$ K in average, in the sense AMBRE minus this study. The mean difference with the effective temperatures estimated from two-micron sky survey \citep[2MASS][]{} colours is found to be equal to $6\pm 100$ K, in the same sense. The final differences with respect to AMBRE were $-0.06\pm 0.30$ dex, and $-0.08\pm 0.20$ dex for log g and the average metallicity, respectively. We point out that such departures are in agreement with the typical external errors reported by the AMBRE Project. Moreover, it is important to note that the reported differences in the parameters can also be explained by the different line lists, analysis procedure and, spectral ranges considered in \citet{wor16} and this study. 

\begin{figure*}
   \centering
   \includegraphics[height=\textwidth,angle=-90]{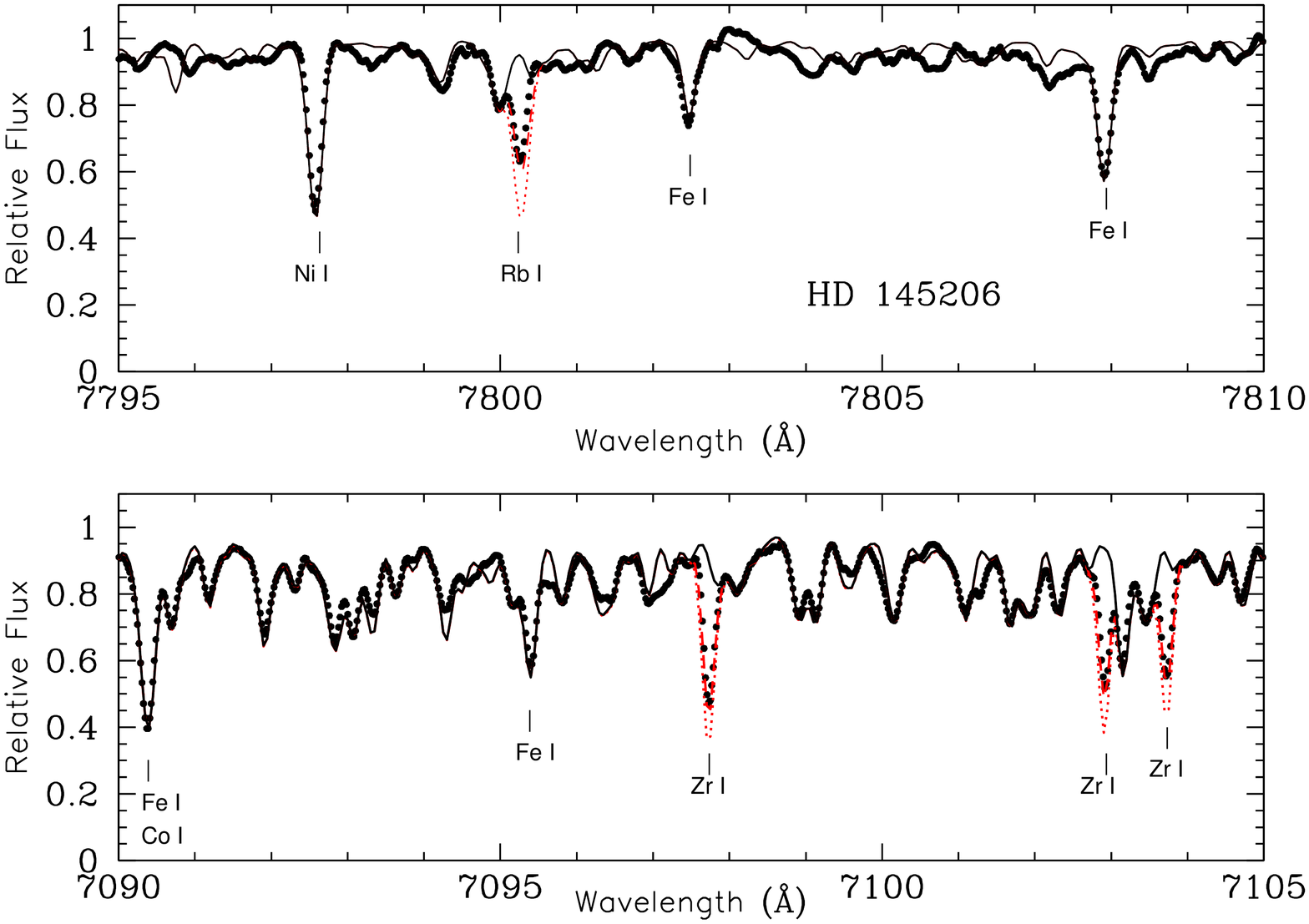}
   \caption{Comparison of the observed (black dots) and synthetic spectra for the M4.0\,III star \object{HD 145206} in the spectral region around $\lambda\lambda 7800\,{\AA}$ (top panel) and $\lambda\lambda 7100\,{\AA}$ (bottom panel). In both panels the continuous black line is a synthetic spectrum with no Rb or Zr, while red dashed and dotted lines show theoretical spectra with log $\epsilon$(Rb)$=2.35$ and 2.6 (top panel), or log $\epsilon$(Zr)$=2.7$ and 3.0 (bottom panel), respectively. Some metallic lines are marked and labelled. In the $\lambda\lambda 7100\,{\AA}$ region the pseudo-continuum is reduced mainly due to the contribution of a TiO veil.} 
   \label{fig:Rb+Zr}
\end{figure*}

The next step was to adjust the determination of the stellar (average) metallicity to our analysis. To do that we used a number of weak metallic lines available in the spectral ranges mentioned above, in particular the \ion{Ti}{i} lines at $\lambda\lambda\sim 7791.34, 7949.15, 8068.23$, and 8069.79 {\AA}; the \ion{Fe}{i} lines at 7095.50, 7802.47, 7807.90, 7941.08, and 7945.84 {\AA}; and the \ion{Ni}{i} lines at 7788.93 and 7797.58 {\AA}. The weakness of these
lines should minimise possible deviations from LTE. In addition, their proximity to
the heavy element (Rb and Zr) lines may reduce systematic effects introduced by the uncertain location of the pseudocontinuum when deriving the elemental ratios with respect to the
average metallicity ([X/M]\footnote{Since for near-solar metallicity stars, it holds that [Ni/Fe]$\approx$[Ti/Fe]$\approx 0.0$, in the following we refer indistinctly to [M/H] or
[Fe/H] as the stellar metallicity.}), since this uncertainty should cancel out. The theoretical fits to these metallic lines also served us to adjust the microturbulence parameter.
In general our metallicities agree within the uncertainty ($\pm 0.15$ dex) with those initially
adopted. Nevertheless, when a difference larger than $0.15$ dex was found, we recalculated a model atmosphere with the new metallicity and repeated the derivation of the metallicity
until convergence was reached. Table 1 summarises the final stellar parameters adopted; last column  indicates the specific bibliographic source for each star. We note that the average metallicity [M/H] and microturbulence velocity shown may not match with the value given in the specific reference as discussed above.

Finally, the abundances of Rb, and Zr were determined by spectral synthesis fits
to the corresponding spectral features. For rubidium, we use the very well known resonance 
lines at $\lambda\lambda 7800$ and 7947 {\AA}, taking into account the hyperfine structure of these lines (see Paper I) and the oscillator strengths from \citet{Morton00}. We adopted the meteoritic $^{85}$Rb$/^{87}$Rb$=2.43$ ratio \citep{lod19}. Unfortunately, the isotopic splitting is tiny and does not allow the derivation of this ratio from our spectra. Concerning zirconium, our main abundance indicator was the \ion{Zr}{i} line at $\lambda 7098$ {\AA} and, as secondary lines those at $\lambda\lambda 7103$, and 7104 {\AA}. In particular these later lines were very useful in the coolest stars of the sample where the \ion{Zr}{i} $\lambda 7098$ {\AA} line may be severely blended with TiO lines. Oscillators strengths for these lines were taken directly from the VALD3 database, with some small corrections after comparison with the observed spectra of Arcturus using the stellar parameters for this star according to \citet{ryd09}. Eventually, we
adopted the solar LTE abundances recommended by \citet{lod19} for Rb (2.47) and Zr (2.58). We also used the solar photospheric abundances recommended by this author for all the other elements. Figure 1 shows an example of  theoretical fits (black and red lines) to the spectral regions of the $\lambda 7800$ {\AA} \ion{Rb}{i} line (top panel) and the \ion{Zr}{i} lines (bottom panel) in a representative star of the sample. Fits to some of the metallic lines used for the determination of the average metallicity are also shown. A small depression of the continuum mainly due to TiO molecule is particularly apparent in the spectral region of the \ion{Zr}{i} lines.

\subsection{Abundance uncertainties and NLTE corrections}
The two main sources of error in the abundances are observational (i.e. related with the SNR of the spectrum) 
and analysis errors caused by the uncertainties in the adopted
model atmosphere parameters. The scatter of the abundances
provided by individual lines of the same species is a good
guide to measurement error. When it was possible ($\sim 70\%$ of the stars), we found excellent agreement
between the Zr abundances derived from the three lines, typically with a
dispersion of less than 0.08 dex. This agreement contrasts with the differences 
($\geq 0.10$ dex) found in the Rb abundance derived from the two lines.
In particular that derived from the \ion{Rb}{i} 7947 {\AA} line is systematically larger by about this
amount. A similar figure was found by \citet{tom99} and \citet{yon06} (the latter in M13 and NGC 6352), which led them
to exclude this line from their analyses in similar stars than here. Thus, we also decide to exclude the $\lambda 7947$ {\AA} \ion{Rb}{i} line from the analysis in this study. We note that a nice agreement between the Rb abundance derived from the two lines was found in Paper I:  typically we found a dispersion of only $\pm 0.02$ dex between the abundances derived from the two Rb lines (see Table 2 in Paper I.) Since the stars studied here are systematically hotter than the M dwarfs in Paper~I, and that effect of telluric lines is very small at the location of this line, we guess that this discrepancy might be caused by an unknown blend with an atomic line with a moderate excitation energy. A detailed study on the formation of these Rb lines
in stars of different spectral types is required to enlight this long standing problem.

The error caused by uncertainties in the adopted stellar parameters can be estimated by
modifying  them by the quoted errors in the analysis of a typical star in the sample and checking the effect on the abundance derived for each species.  To do this we have adopted the uncertainties estimated in the AMBRE project \citep{wor16}, since for most of the
stars we adopted the stellar parameters  derived in this survey (see Table 1), namely: $\pm 100$ K in $T_{\rm eff}$, $\pm 0.2$ dex in log g, $\pm 0.2$ km s$^{-1}$ in $\xi$, $\pm 5\%$ in C/O, and $\pm 0.15$ dex in [Fe/H]. For a typical giant star in the sample with parameters $T_{\rm eff}/$log $g/$[Fe/H]$=4050/1.5/0.0$, we found that the abundances
derived are mostly affected by the uncertainty in $T_{\rm eff}$: $\pm0.07$ and $\pm 0.14$ dex for Rb, and Zr, respectively.  Uncertainties in the gravity, metallicity, microtubulence and the C/O ratio are relatively low for Rb, namely: $\mp 0.03, \mp 0.05, \mp 0.05$ and $\mp 0.04$ dex, respectively, while they are rather significant for Zr: $\mp 0.10$ and $\mp 0.15$ dex for gravity and metallicity, respectively. However, the above quoted uncertainties in the microturbulence and C/O have almost no effect on the Zr abundance. Adding these uncertainties together quadratically, we estimated a total uncertainty in [X/H] of $\pm 0.15$ dex for Rb, and $\pm 0.23$ dex for Zr.  These estimates include the continuum location uncertainty (about 1-2\%) as an independent source of error and, in the case of Zr, the typical dispersion ($\pm 0.04$ dex) around the mean abundance value when more than one line was used.
Nevertheless, the abundance of these elements relative to average metallicity, [X/Fe], holds the most interest. This ratio is more or less sensitive to the uncertainties in the atmospheric parameters depending on
whether changes in the stellar parameters affect the heavy element abundance and metallicity in the same or opposite sense. In our case, we estimated total uncertainties of $\pm 0.12$ dex and
$\pm 0.20$ dex for the [Rb/Fe] and [Zr/Fe] ratios, respectively. Certainly the internal (relative) errors within the sample studied would be smaller.

On the other hand, the structure of the atom of Rb is very similar to that of other alkaline elements, such as Na and K. It is very well known that the resonant lines of these alkaline elements are affected by deviations from LTE \citep{Bruls92}. Recently \citet{Korotin20} (see also Paper I) has estimated the LTE deviations in the formation of the Rb lines as a function of the effective temperature, gravity, metallicity, microturbulence and [Rb/Fe] ratios in dwarf and giant stars. This study shows that the NLTE corrections (in the sense $\Delta_{\rm{NLTE}}=$N$_{\rm{NLTE}}-$N$_{\rm{LTE}}$ abundances) vary in a non trivial way depending on the stellar parameters. Here, we have estimated NLTE corrections to the Rb abundances derived from the $\lambda 7800$ {\AA} line star by star according to \citet{Korotin20}. The corrections are shown in Table 1 (column seven) where it can be seen that they can be positive or negative depending on the stellar parameters, and may reach up to $-0.15$ dex for stars with $T_{\rm eff}\sim 4000$~K and [Rb/Fe]$>0.0$\footnote{In Paper I we showed that the average NLTE Rb abundance corrections in M dwarfs is $\sim -0.15$ dex. However, the [Rb/Fe] ratios derived remain equal respect to the LTE analysis since the NLTE corrections are almost
compensated by the Solar NLTE abundance, which is 0.12 dex (2.35) lower than the LTE value (2.47).}. We refer to \citet{Korotin20} for a detailed discussion on this topic. It is worth noting that the Solar NLTE Rb abundance found in this paper is 2.35, which is in excellent agreement with that found in meteorites \citep{lod19}.
Unfortunately there is a very limited information in
the literature concerning the NLTE corrections for the LTE Zr abundance derived from different lines, although it appears that for solar metallicity stars they may be small \citep[see][]{vel10}.
   \begin{figure}
   \centering
    \includegraphics[width= 9.5 cm]{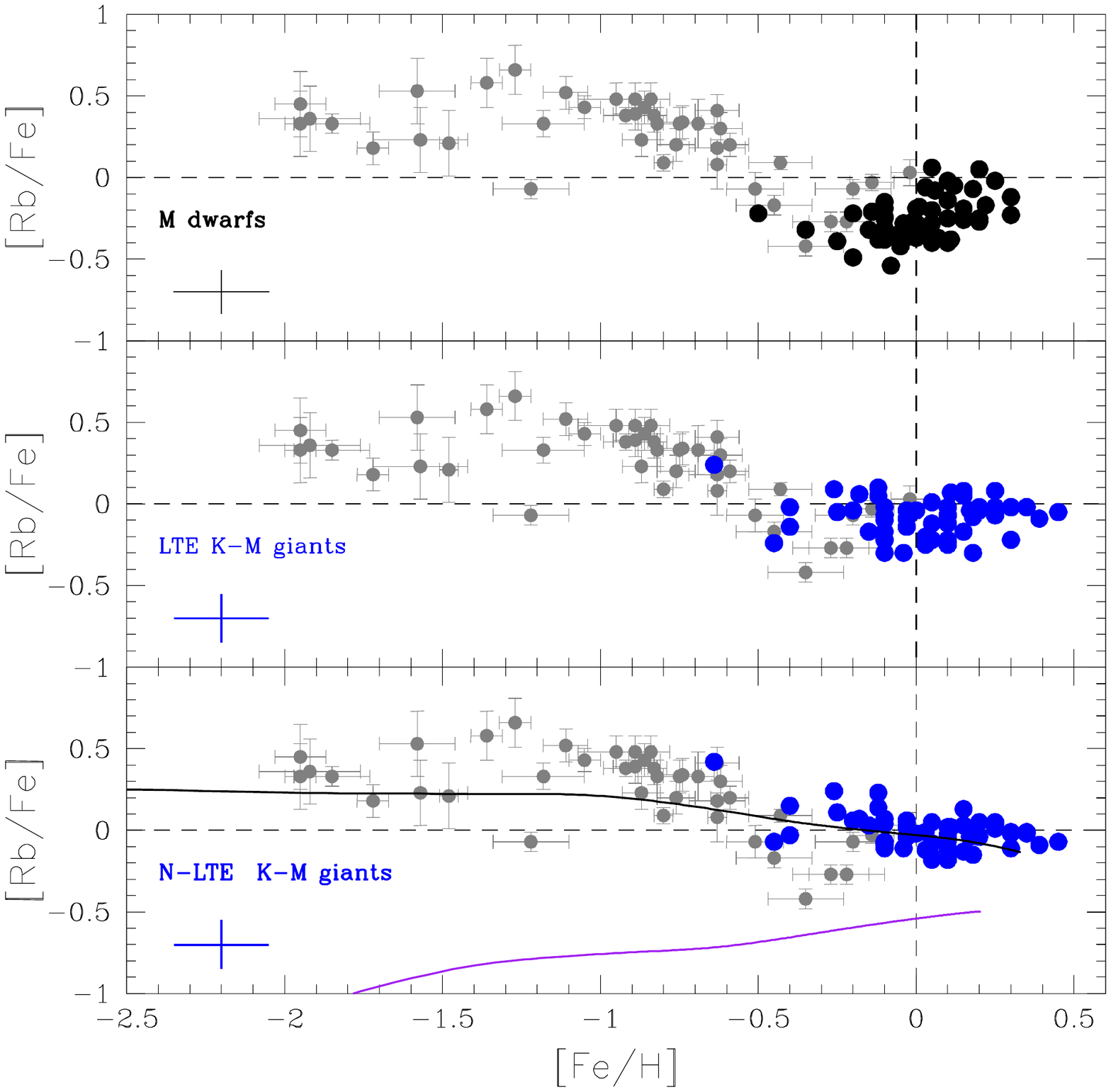}
    \caption{[Rb/Fe] vs. [Fe/H] diagram for M-dwarfs studied in Paper I (top panel, black dots, LTE abundances) and for KM-type giants (blue dots) in this study in LTE (middle panel) and NLTE (bottom panel). The grey dots with error bars in the three panels are the [Rb/Fe] ratios derived in halo and disc giant and dwarf stars by \citet{gra94} and \citet{tom99}. A typical error bar in the [Rb/Fe] ratios in Paper I and this study is shown in the bottom left corner of each panel. Upper limits in the Rb abundances are omitted in the figure. In the bottom panel, solid curves are theoretical GCE predictions by \citet{pra18,pra20}: Black line includes the contributions from LIMS and RM stars, and the r-process, while magenta  line include only LIMS (see text for details). }
    \label{fig:[Rb/Fe]}
\end{figure}

\section{Results and discussion}
   
  Table 1 shows the final Rb and Zr abundances derived in our stars. When more than one Zr line was used the abundance quoted is the average value. 
  
  Figure 2 shows the observed [Rb/Fe] vs. [Fe/H] relationship obtained in our stars (middle and bottom panels, blue dots) compared with that obtained in M dwarfs in Paper I (top panel, black dots). In the three panels we have also included the [Rb/Fe] ratios obtained by \citet{gra94} and \citet{tom99} (grey dots) in metal-poor GK dwarfs and giants\footnote{The [Rb/Fe] ratios in these studies have been scaled to the Solar LTE Rb abundance adopted here. For the typical atmosphere parameters in their stellar sample, NLTE corrections for these stars are within 0.07-0.10 dex (negative), however the corresponding [Rb/Fe] ratios would change only slightly due again to the lower NLTE Solar Rb abundance.}. First, focusing on the LTE Rb abundances (middle panel), it is apparent that the [Rb/Fe] vs. [Fe/H] relation derived  is nearly flat in the full metallicity range studied (excluding the moderate metal-poor giant HD 1638, see Table 1), showing a small deficiency with respect to the solar value: average [Rb/Fe]$=-0.07\pm 0.11$~dex, which is compatible with [Rb/Fe]$\approx 0.0$ within the error bar. Furthermore, no trend with  the increasing metallicity is seen. This clearly contrasts with the relationship obtained for M dwarfs in Paper I  where a systematic deficiency  by a factor two (on average) with respect to the Sun and a hint of increasing [Rb/Fe] with metallicity were obtained (see top panel in Fig. 2). This is even more evident when NLTE Rb abundances are considered (bottom panel,  blue dots). Furthermore, in this case  the dispersion of the [Rb/Fe] at a given [Fe/H] diminishes significantly around the solar value
(average [Rb/Fe]~$=-0.01\pm 0.09$ dex). As a consequence the [Rb/Fe] ratio behaves very similarly to that observed for [Eu/Fe]  \citep{bat16,del17,for19} -- Eu being an almost pure r-process element -- at least up to [Fe/H]~$\sim -2.0$ dex: i.e. a nearly constant [Rb/Fe] ratio for [Fe/H]~$\lesssim -1.0$ dex and then a smooth decrease to reach [Rb/Fe]~$\approx 0.0$ at solar metallicity. We note that the [Rb/Fe] vs. [Fe/H] seems totally flat at [Fe/H]~$>0.0$, however the [Eu/Fe] ratios apparently become negative for metallicity larger than solar \citep[see e.g.][]{bat16}. This would imply that at this metallicity Rb has different contributing sources than Eu; more Rb abundance determinations at metallicity larger than solar are needed to confirm this figure.

Figure 2 (bottom panel) compares the observed relationship with model predictions  from a 
galactic chemical evolution (GCE) model of \citet{pra18,pra20}, which includes Rb contributions from low-and-intermediate mass stars (LIMS), rotating massive stars (RMS) and the r-process (black continuous line). Note taht the GCE model  from \citet{pra18} used here is a one-zone model tailored for the solar neighborhood. It is meant to reproduce the evolution of the chemical composition of the local gas, reaching a final metallicity (at age 0 Gy) of [Fe/H]$\sim +0.1$, slightly above the metallicity of the local gas and of the youngest stars e.g. in Orion. Therefore, is not well suited
 to deal with the super-metallicity stars currently found in the solar neighbourhood, as those shown in Figs. 2 and 3.  It is widely accepted that stars of supersolar metallicities locally observed are attributed  to radial migration: they are formed in the inner disk (which had a different  chemical evolution history and reached supersolar metallicities early on) and they have migrated at $\sim 8$ kpc. As a consequence, they do not have to be young many of them could have ages as large as $\sim 4$ Gy. Several models deal with those issues \citep[see e.g.][]{min13,kub15}. However, our 1-zone model used here is sufficient for our discussion, as far as we do not enter the super-solar metallicity regime\footnote{A note of caution here: if some nucleosynthesis effect depends on metallicity (e.g. secondary elements from s-process) then the effect should show up clearly as function of metallicity, independently of the stellar age.}. According to this GCE model, the production of Rb through the weak s-process in RMS is critical to account for the observed relationship, in particular at  solar metallicity. Note that the contribution only by LIMS  through the main s-process (magenta line in Fig. 2) is clearly not enough, as expected according the $\sim50$\% s- and r-process origin for the bulk Rb abundance observed in the Solar System (see \citet{pra18,pra20} for a detailed discussion on the stellar yields adopted). The [Rb/Fe] vs. [Fe/H] relationship obtained is now nicely reproduced without invoking any non-standard nucleosynthesis process for Rb for metallicities close to solar in contrast to our suggestion in Paper I.

 \begin{figure}
   \centering
    \includegraphics[width= 9.5 cm]{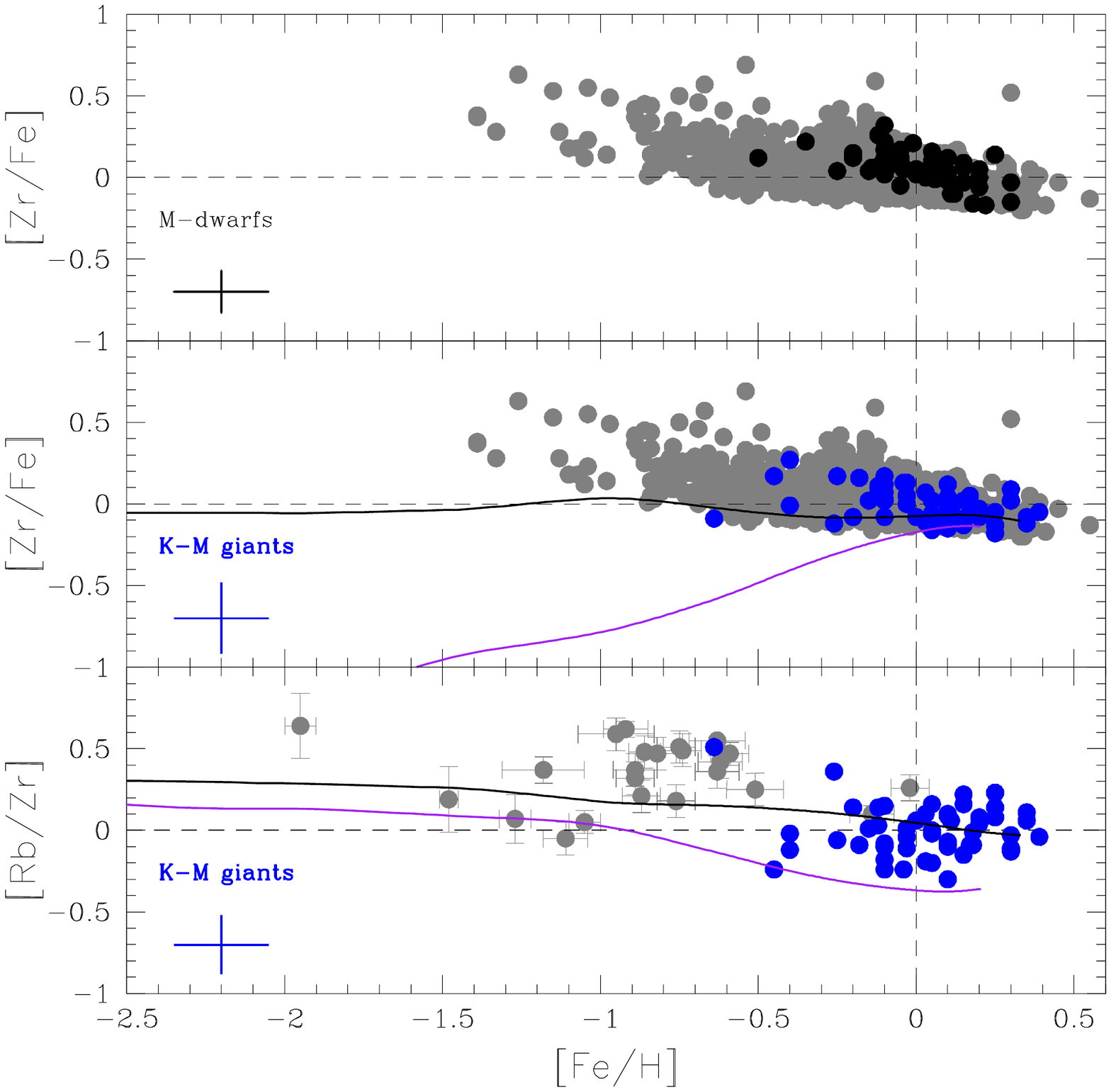}
    \caption{Top and middle panels: same as Fig. 2 but for LTE [Zr/Fe] vs. [Fe/H] for M dwarfs studied in Paper I (top panel, black dots) and for KM-type giants (blue dots) in this study (middle panel). Grey dots are the [Zr/Fe] ratios derived by \citet{bat16} and \citet{del17}, both in thin and thick disc dwarf stars. Bottom panel: [Rb/Zr] vs. [Fe/H] for the stars in this study (blue dots) when using NLTE Rb abundances. Grey dots correspond to the giants and dwarfs stars analysed in common by \citet{gra94}, \citet{tom99}, and \citet{mis19}. A typical error bar in the abundance ratios is shown in the bottom left corner of each panel. For the data in the literature (grey dots) the error bars have been omitted for clarity. Upper limits in the Zr abundance have been also omitted. Continuous solid lines in middle and bottom panels are the GCE predictions as in Fig. 2.}
    \label{fig:[Zr/Fe]}

\end{figure}

 Figure 3 shows the [Zr/Fe] vs. [Fe/H] relationship derived in our stars (middle panel, blue dots) compared with the relation obtained in the most recent similar analyses in GK dwarfs by \citet{bat16}  and \citet{del17} (grey dots in top and middle panels). We compare also with the results obtained in Paper I for Zr (top panel, black dots) in M dwarfs. From this figure, it is evident that the [Zr/Fe] vs [Fe/H]  behaviour obtained here is almost identical to that for M dwarfs in the metallicity range studied, and both indistinguishable from that derived in GK dwarfs. We note a slight tendency for [Zr/Fe] to decrease with metallicity, even for [Fe/H]$>0$ as previously found \citep{bat16,del17,for19}. This supports the Zr abundances derived in M dwarfs in Paper I, but puts some doubts on the reliability of Rb abundances obtained there. 
 
 On the other hand, the observed [Zr/Fe] vs. [Fe/H] relationship is also nicely accounted by the GCE model of \citet{pra18} when all the contributing sources for Zr are considered (black curve in Fig. 3). Again LIMS are not sufficient (magenta curve in Fig. 3) to adjust the observed trend, although their contribution at [Fe/H]$\sim 0.0$ is  more important in the case of Zr than of Rb. This agrees with the $\sim 82$\% s-process contribution to the abundance of this element in the Solar System \citep[see e.g.][]{pra20}. 
 Recently it has been reported a possible increasing trend of the [Zr/Fe] ratio (and of other s-elements; Y, Ba, La, Ce) with age in  supersolar metallicity stars belonging to young open cluster \citep[see e.g.][]{mai12,mis15,mag18}.  Our GCE predictions cannot reproduce this apparent increase of the [Zr/Fe] ratio at very young age. It has been argue that the i-process and/or a non standard s-process nucleosythesis in low-mass AGB stars, might explain this observational trend (see references in the studies above). In any case, if we plot  [Zr/Fe] vs. time according to our GCE model, we obtain an almost flat curve from 12.5 to 0 Gy age
 around [Zr/Fe]$\sim 0.0$, which is fully compatible with the observational results by \citet{mag18} (see their Fig. 9) obtained for the stars belonging to the thin disk (from their kinematic properties, we deduce that the overwhelming majority of the stars studied here belong to the thin disk). However, this apparent increase of the [X/Fe] ratios
 of s-elements in young open cluster is still rather controversial: at least, as far as the [Ba/Fe] ratio concerns, this trend has  been shown to be correlated with the stellar activity of young stars and to not be nucleosynthetic in origin \citep[see][]{red17}, putting serious doubts on the reliability of this increasing trend with age. In fact, we already addressed this issue in Paper I (at the end of Section 3) when discussing the observed trend of increasing [Rb/Fe] vs. [Fe/H] in
metal-rich stars, trend which we discard now in this study. 
 
Finally, the bottom panel in Figure 3 shows the [Rb/Zr] ratios derived here against [Fe/H]. This figure should be compared with the equivalent Fig. 8 in Paper I for M dwarfs. Similarly to that figure in Paper I, Fig. 3 shows that as metallicity
increases, the [Rb/Zr] diminishes and cluster around [Rb/Zr]$\sim 0.0$ 
on average for [Fe/H]$\sim 0.0$, although with a much less dispersion than that obtained in Paper I. This  dispersion is consistent with the uncertainties in the present analysis. The
decrease in the [Rb/Zr] ratio for increasing metallicity is clearly due to the increasing relevance of the contribution of low-mass stars in the production of Rb and Zr through the main
s-process, for which the $^{13}$C$(\alpha,n)^{16}$O is the main neutron source. When this neutron source is at work, [Rb/Zr]$ < 0.0$ is expected at metallicities close to solar \citep[see e.g.][]{lam95,abi01}, as shown by the GCE model (magenta line). However, when the Rb production in RMS is included in the GCE model, the full observed (average) relationship can be nicely reproduced as shown in Fig. 3 (black line). Therefore, from Figs. 2 and 3 one can conclude that the observed evolution of the [Rb,Zr/Fe] ratios at metallicity close to solar, can be understood within our current understanding of the Rb and Zr production in rotating massive and low-and-intermediate mas stars through the weak and main s-process nucleosynthesis, respectively.


 
   \subsection{Rubidium deficiency in M dwarfs}
Why M dwarfs with near solar metallicity apparently show Rb deficiencies with respect to the Solar value? Is this finding  real? In Paper I we discussed  various issues that might account for this (NLTE effects, stellar activity, or an anomalous Rb abundance in the Solar System), but no  satisfactory explanation was found. Here we address again this issue discussing the impact of gravitational settling and the existence of a magnetic field of moderate intensity in the surface of M dwarfs.

\subsubsection {Gravitational settling}
Comparison of evolutionary tracks \citep{bre12,tan14,bar15} for M dwarfs of $0.4-0.7$ M$_\odot$ and $Z\sim Z_\odot$ in the $T_{\rm eff}-$ log $g$ diagram shows that the M dwarfs studied in Paper I have ages 
$\sim 5$ Gyr in average \citep[see][]{pass18}. Probably, there are older M dwarfs within this sample, but evolutionary tracks for different masses and metallicities become rather degenerate in age for ages larger than $\sim 2$ Gyr. This fact, together with the $T_{\rm eff}$ and log $g$ uncertainties, make difficult an accurate age estimate. In any case, and depending on the specific stellar mass and metallicity, for such old ages the surface chemical composition  of  these stars may have been altered due to gravitational settling, with differential depletion for some metals. In fact, it is very well known that the current surface chemical composition of the Sun is different from that in the proto-solar nebulae 4.56 Gyr ago \citep{lod03}. In particular, \citet{pie07} showed that the proto-solar Rb abundance was a $\sim 8$\% higher than the present Solar photospheric value due to the operation of gravitational settling. For stellar masses lower than the Sun and Z$\sim Z_\odot$, larger surface Rb depletion would occur at an age $\sim 5$ Gyr. However, this depletion would affect in a similar way to the  neighbouring elements Sr, Y and Zr, thus no relative effect between Rb and these elements would be expected\footnote{Note that the isotope $^{87}$Rb would be depleted in a larger factor ($\sim 20\%$) because its radiative decay (t$_{1/2}\sim 4.92\times 
10^{10}$ yr) into $^{87}$Sr. However, the abundance of this isotope represents only $\sim 27\%$ of the total Rb abundance.}. This is at odd with that observed in Paper I where no hint for any Sr and/or Zr depletion was found. Furthermore, we note that stars in the mass range $0.4-0.6$ M$_\odot$  have about $\sim 0.1$ M$_\odot$ in their convective envelope near the turn-off at solar metallicity. This acts as a good buffer to changes in the surface composition  keeping everything near the surface nicely stirred up: i.e. the thickness of the convective envelope of a typical M dwarf at solar metallicity should prevent any changes by gravitational settling from being significant. One has to conclude, therefore, that gravitational settling  is probably not the cause of the observed Rb deficiency in M dwarfs.

 \begin{figure*}
   \centering
    \includegraphics[height=\textwidth,angle=-90]{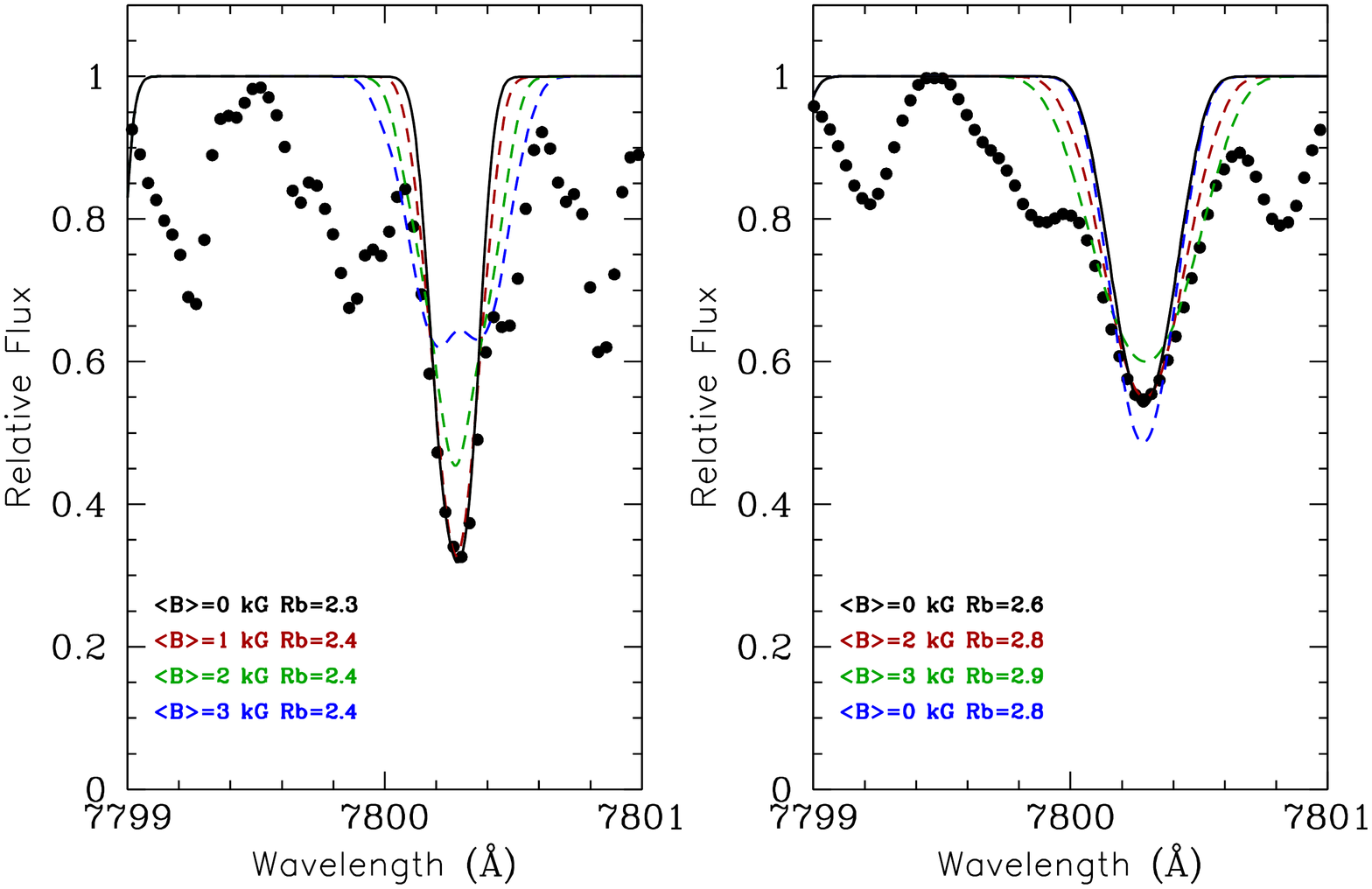}
    \caption{Comparison of the observed spectrum (black dots) at the location of the $\lambda 7800$ {\AA} of \ion{Rb}{i} line  for the M dwarfs G244-77 (non-active, left panel) and OT Ser (active, right panel) studied in Paper I, with synthetic spectra (continuous and dashed lines) computed with different average magnetic field intensities in the line of sight and Rb abundances (as labelled). Only the \ion{Rb}{i} line is included in the spectral synthesis (see text for details).}
    \label{fig:[RbB]}

\end{figure*}
  
   \subsubsection{Zeeman broadening}
    
On the other hand, in Paper I we qualitatively discussed  the effect of  magnetic fields in the profile of the spectral lines with large Land\'e factors (as the Rb resonance lines) in active M dwarfs. The affected lines appear shallower and broader, and this may affect the abundance determination. In fact, in Paper I we identified three stars (J11201–104, OT Ser, and J18174+483), with strong activity levels based on their H$_\alpha$-emission as a proxy for activity indicator. For instance, \citet{sch19} derived an average field intensity $\langle B \rangle \sim 3$ kG in OT Ser. By comparison with non-active stars of the same spectral type, we confirmed that in these three stars the Rb lines were indeed affected by Zeeman broadening, and that the Rb abundances derived were amongst the lowest derived ([Rb/Fe]$<-0.30$ dex) in the stellar sample. Here we address this issue more quantitatively. 

The existence of strong magnetic fields in M dwarfs is known since 
\citet{saa85} based on the analysis of high resolution spectroscopy infrared spectra.
Several recent studies report field strength measurements in the range from 0.8 to 
7.3~kG \citep[see e.g.][]{shu19}, although the spectroscopic requirements for detecting subtle signatures
of the Zeeman broadening may be satisfied for only a small number of the brightest active M dwarfs. Furthermore, interpretation of these signatures became ambiguous as soon as the stellar rotational velocity exceeded $\sim 5$ km s$^{-1}$ since line profile details are washed out by the rotational Doppler broadening. This made it impossible to probe magnetic fields in faster rotating and, presumably, most magnetically active M dwarfs.
A detailed discussion on the effects of magnetic fields in the spectrum of  M dwarfs is obviously beyond the scope of this study. Our aim here is only to quantify how the Rb abundance derived may be affected by the presence of an average magnetic field in a typical M dwarf, even in those which are considered as {\it non  active} where no Zeeman broadening is seen on the profile of the spectroscopic lines that potentially could be affected. An excellent review on magnetic fields in M dwarfs together with the observational techniques used for its determination can be found in \citet{koc20}.

The synthesis of the \ion{Rb}{i} lines under the presence of a magnetic field needs
to be done under the intermediate Paschen-Back effect using the proper hyperfine and
Zeeman effective Hamiltonians \citep[see, e.g.,][]{2007ApJ...659..829A}. The proximity
of the hyperfine energy levels produces interferences among the magnetic sublevels
when a magnetic field is present, so that one needs to resort to the diagonalisation
of the full Hamiltonian for computing the wavelength shift of every magnetic component
on the Zeeman pattern. The synthesis is done by solving the polarised radiative transfer
equation for the Stokes vector $(I,Q,U,V)$ using the DELOPAR method \citep{2003ASPC..288..551T}. The emission vector and the propagation matrix is computed under the assumption of LTE using the expressions found in Section 9.1 of \cite{2004ASSL..307.....L} for different orientations of the magnetic field vector. Rubidium contains two main isotopes with non-negligible abundance (see above). Both isotopes have different nuclear spins ($^{87}$Rb has $I=3/2$, while $^{85}$Rb has $I=5/2$) and the isotopic shifts of their energy levels is smaller than the width of the line, so that both need to be considered as blends when computing the opacities. The line at $\lambda 7800$~\AA\ of interest here is a hyperfine multiplet produced by the transition $^2S_{1/2}-^2P_{3/2}$.
The isotopic shifts $\delta=E_{85}-E_{87}$, i.e., the
energy difference for a given level between the level for $^{85}$Rb and $^{87}$Rb are:
$\delta(^2 \mathrm{S}_{1/2})=164.35$ MHz, and
$\delta(^2 \mathrm{P}_{3/2})=86.31$ MHz, respectively \citep{2011PhRvA..84c4501A}.
The hyperfine effective Hamiltonian was parametrised in terms of the magnetic-dipole and 
electric-quadrupole hyperfine structure constants. Since the effect of the electric-quadrupole constant is almost negligible, we only use the magnetic-dipole constant. We adopted the following values for $^{87}$Rb: $A(^2S_{1/2})=3417.341$ MHz, $A(^2P_{1/2})=406.2$ MHz and $A(^2P_{3/2})=84.845$ MHz, obtained from \cite{PhysRevA.83.052508}. For $^{85}$Rb we use: $A(^2S_{1/2})=1011.910$ MHz, $A(^2P_{1/2})=120.721$ MHz and $A(^2P_{3/2})=25.0091$ MHz, obtained from \cite{RevModPhys.49.31}. Finally, the Land\'e factor of each fine structure level is obtained by assuming LS coupling, which gives $g(^2S_{1/2})=2$, $g(^2P_{1/2})=2/3$ and $g(^2P_{3/2})=1.3362$.

Figure 4 compares the observed spectrum of a non active M dwarf (G244-77, left panel) and an active one (OT Ser, right panel) \citep[see][]{sch19} - both stars studied in Paper I - at the position of the \ion{Rb}{i} $\lambda 7800$ {\AA} line, with synthetic spectra computed considering different Rb abundances and average magnetic field intensities in the line of sight computed as described above. Only the \ion{Rb}{i} line has been included in the spectral synthesis. For the active star OT Ser (right panel), synthetic spectra were convolved with a rotational profile (together with the instrumental profile) since this star shows $v\sin{i} > 2$ km s$^{-1}$ \citep{Rei18}. From Figure~4 it can be clearly appreciated that the profile of the \ion{Rb}{i} line in OT Ser is much broader and shallower than in the no-active star G244-77 (left panel), considering that both stars have very similar stellar parameters (see Table 1 in Paper I). This is due to the combined effect of rotation and a strong magnetic field in OT Ser. For a given Rb abundance (2.6) and assuming an average field $\langle B \rangle=0$, the computed synthetic spectrum clearly fails to fit the wings of the line despite it does fit the core; while when an average field similar to that observed ($\sim 2-3$ kG, see above) is included, the fit improves considerably. What matters here is the difference in the Rb abundance between both cases: from the figure it comes off that this difference may be up to $0.2-0.3$ dex, in the sense of higher Rb abundance when magnetic field is included in the synthesis. We note that this value is roughly the average systematic Rb deficiency with respect to the solar value found in M dwarfs in Paper I. Interesting enough, Figure 4 (left panel) shows that
even for a non-active M dwarf as \mbox{G244-77}, considering a {\it weak} magnetic field ($\sim 1$ kG) in the spectral synthesis may imply a Rb abundance difference up to $\sim 0.1$ dex higher with respect to the case with no magnetic field. We note that the observational threshold for the detection of an average magnetic field in M dwarfs is roughly 1 kG \citep[see e.g.][]{koc20}. Therefore, even in M dwarfs considered as non-active with non apparent Zeeman broadening in the spectrum, the resonance Rb I lines may be indeed affected by an average weak magnetic field and, as a consequence, Rb abundances may be underestimated. The exact amount of this effect would depend on the orientation of the average magnetic field along the line of sight, which is rather difficult to discern observationally. Fig. 4  shows the case of maximum effect occurring when the magnetic field is parallel to the line of sight: the larger inclinations, the deeper becomes the core of the 7800 {\AA} line so that the abundance difference with respect to the absence of magnetic field is reduced correspondingly. Then, for an expected uniform distribution of magnetic field inclinations respect to the line of sight in a given sample of observed M dwarfs with different levels of activity, one would expect a uniform distribution of Rb abundance corrections, which would be within the range $\sim 0.0-0.3$ dex (increase) in the case of the M dwarfs studied in Paper I. This would considerably alleviate the difficulty to explain the Rb deficiency found, although would not fully solve the issue. Obviously, magnetic fields also exist at the surface of KM-type giants \citep[see e.g.][]{aur15} but their intensity is much smaller (a few tenths of Gauss) than those observed in M dwarfs and, therefore, its effect would be negligible.

\section{Conclusions}
We have compiled a sample of 54 bright K- and M-type giant stars with metallicity close to solar. New Rb and Zr abundances are derived from the high-resolution, high-SNR spectra parametrised by \citet{wor16} within the AMBRE Project. Our aim is to test the reliability of the Rb deficiency recently found in a sample of M dwarfs
in a similar metallicity range by \citet{abi20} (Paper I). Based on the
observational data analysed, our main conclusions can be summarised as follows.

1. The LTE [Rb/Fe] ratios derived in our sample stars show in average a slight deficiency with respect to the solar value, [Rb/Fe]$\approx -0.07\pm 0.11$ dex, nevertheless smaller than that found in Paper I.
However, when a NLTE analysis  is done, this deficiency disappears and the [Rb/Fe] ratio clusters around the solar value with a small dispersion, [Rb/Fe]$\approx -0.01\pm 0.09$ dex. This contrasts with the results in M dwarfs for Rb in Paper I.  

2. The [Zr/Fe] ratios derived are very similar to the most recent determinations in FGK dwarfs of similar metallicity, which support our analysis for Rb. 

3. As a consequence, the [Rb/Fe] and [Rb/Zr] vs. [Fe/H] relationships obtained in the metallicity range studied can be explained through a chemical evolution model for the solar neighbourhood when the Rb production by rotating massive stars and low-and-intermediate mass (AGB) stars (these later also producing Zr), are  considered according to the yields from \citet{lim18} and \citet{cri15}, respectively, without the need of any deviation from the standard s-process nucleosynthesis in AGB stars, contrarily to what was suggested in Paper I.

4. We explore if gravitational settling and magnetic activity may be the cause of the
Rb deficiency previously reported for M dwarfs. While the first phenomenon would have little impact on the surface Rb abundance in these stars, we show that when the Zeeman broadening is included in the spectral synthesis for the typical average magnetic field intensity observed in M dwarfs, the Rb abundances derived may increase  significantly. This can explain, but not totally, the discrepancy between the Rb abundances derived in solar metallicity M dwarfs and KM-type giants.

We conclude then that, although  abundance  analysis  in M dwarfs well illustrates its  value for Galactic  chemical evolution  studies, attention has to be paid  when deriving elemental abundances from spectral atomic lines formed in the upper layers of their atmospheres, whether affected or not by magnetic activity. This will be important, for instance, for future spectroscopic follow-up observations of the PLATO mission, among others. More generally, the complexity of the physical processes influencing Rb abundance estimates illustrated here shows the importance of carefully considering all the stellar physical properties in any spectral analysis. This is particularly true for large scale surveys dealing with a variety of stellar types.

\begin{table*}
\caption{Stellar parameters and abundances derived in the sample of stars$^a$.}         
\label{par_ab_data}      
\centering          
\begin{tabular}{lccccccccc}     
\hline
\hline       
\noalign{\smallskip}
Star & T$_{\rm {eff}}$ (K) & log ${g}$ & [M/H]& $\xi$ (km s$^{-1}$) & C/O& $\log{\epsilon({\rm Rb})}_{\rm{LTE}}$  & $\Delta_{\rm{NLTE}}$ (dex) & $\log{\epsilon({\rm Zr})}$& Reference$^b$ \\ 
\hline                    
\noalign{\smallskip}
\object{HD 1638} & 4138& 1.25& $-0.64$& 1.8&0.36 &2.07&   0.06& 1.85& 1\\
\object{HD 5544} & 4443& 2.20&    $0.05$& 1.5 &0.61 & 2.30&$-$0.06& 2.47& 1 \\
\object{HD 11643}& 4412& 2.09&    0.25& 1.6& 0.59 &2.70&$-$0.09& 2.70& 1\\
\object{HD 12642}& 3826& 0.78&    0.11& 2.0&0.60  &2.65&$-$0.17& 2.65&  1\\
\object{HD 17361}& 4477& 2.34&    0.10& 1.8& 0.43& 2.50&$-$0.03& 2.60&  1\\
\object{HD 18884}& 3796& 0.68& $-0.45$& 1.8&0.53 &1.78&   0.05& 2.30& 2\\
\object{HD 29139}& 3814& 1.00& $-0.03$& 1.9&0.59 &2.30&$-$0.02& 2.55& 1 \\
\object{HD 31421}& 4440& 2.13& $-0.10$& 1.5& 0.55&2.27&$-$0.09& 2.65& 1\\
\object{HD 61603}& 3809& 1.04&    0.09& 1.8& 0.57&2.50&$-$0.16& 2.65&  1\\
\object{HD 65354}& 3903& 0.51& $-0.10$& 1.7&0.54 &2.20&$-$0.02& 2.51& 1\\
\object{HD 71160}& 4100& 1.70&    0.30& 1.8& 0.61&2.75&$-$0.12& 2.97& 3\\
\object{HD 72505}& 4553& 2.50&    0.25& 1.8& 0.65&2.65&$-$0.02& 2.78& 4\\
\object{HD 74088}& 4020& 1.69& $-0.26$& 1.8&0.59 &2.30&   0.03& 2.20&  3\\
\object{HD 78479}& 4418& 2.20&    0.35& 1.7& 0.54&2.80&$-$0.11& 2.81&  4\\
\object{HD 79349}& 3884& 1.79&    0.15& 1.8&0.52 &2.67&$-$0.14& 2.60&  5\\
\object{HD 81797}& 3977& 1.14&    0.03& 1.6& 0.63&2.30&$-$0.04& 2.68& 1\\
\object{HD 83240}& 4400& 2.47&    0.10& 1.6&0.52 &2.45&$-$0.06& 2.53& 1 \\
\object{HD 90862}& 3899& 1.07& $-0.40$& 1.7& 0.54&1.93&$-$0.01& 2.17& 1\\
\object{HD 93813}& 4310& 1.87&    0.05& 1.6& 0.58&2.40&$-$0.08& 2.57& 6\\
\object{HD 95208}& 4131& 1.16& $-0.04$& 1.4&0.59 &2.13&   0.07& 2.67& 5\\
\object{HD 95849}& 4472& 1.17&    0.18& 2.1& 0.63&2.57&   0.02& 2.74& 4\\
\object{HD 102212}& 3812& 0.86&$-0.10$&  2.0&0.30 & 2.15&   0.01& ...   & 1\\
\object{HD 102780}& 3900& 1.60&$-0.11$&  1.6&0.58 &2.07&   0.07& 2.55& 3\\
\object{HD 107446}& 4100& 1.24&  0.10&  1.5&0.55 &2.55&$-$0.13& 2.72& 1\\
\object{HD 111464}& 4160& 1.36&  0.15&  1.6& 0.58&2.45&$-$0.08& 2.75& 4\\
\object{HD 119971}& 4093& 1.36&$-0.40$& 1.8& 0.49&2.05&   0.05& 2.45& 7\\
\object{HD 121416}& 4576& 2.07&   0.35& 1.7&0.53 &2.80&$-$0.12& 2.85& 1\\
\object{HD 124186}& 4290& 2.50&   0.45& 1.5&0.54 &2.87&$-$0.14& ...   & 1\\
\object{HD 128688}& 4083& 1.30&   0.17& 1.4&0.50 &2.60&$-$0.12& 2.80& 1\\
\object{HD 132345}& 4400& 2.49&   0.39& 1.7& 0.55&2.77&$-$0.12& 2.92& 1\\
\object{HD 138716}& 4700& 2.50&   0.20& 1.7& 0.47&2.63&$-$0.03& 2.75& 1\\
\object{HD 140573}& 4540& 2.50&   0.30& 1.8&0.57 &2.55&$-$0.01& 2.90& 6\\
\object{HD 143107}& 4283& 1.93&$-0.03$& 1.7& 0.54&2.40&$-$0.02& 2.60& 1\\
\object{HD 145206}& 4020& 1.43&   0.18& 1.6&0.67&2.35&   0.03& 2.70& 1\\
\object{HD 146051}& 3850& 1.20&   0.10& 1.9& 0.53&2.35&$-$0.08&$<$2.60& 1\\
\object{HD 148291}& 4545& 1.64&$-0.10$& 1.6& 0.49&$<$2.30&$-$0.01&2.60& 8\\
\object{HD 148513}& 4000& 0.80&   0.20& 1.7& 0.67&2.65&$-$0.14& 2.70&  6\\
\object{HD 149161}& 3848& 1.00&$-0.21$& 1.9& 0.53&2.23&$-$0.02& 2.30& 1\\
\object{HD 149447}& 3808& 0.72&$-0.12$& 1.9&0.42 &2.40&$-$0.03& 2.57& 1\\
\object{HD 152786}& 3813& 0.13&$-0.15$& 2.0& 0.57&2.15&   0.08& 2.45& 1\\
\object{HD 157244}& 4233& 1.17\   &0.02& 2.2& 0.58&2.25&   0.12& 2.50& 1\\
\object{HD 167006}& 3600& 1.00&$-0.10$&  2.0&0.53 &2.15&   0.02&$<$2.60& 3\\
\object{HD 167818}& 4001& 0.50&$-0.25$& 1.7&0.58 &2.17&   0.04& 2.50& 4\\
\object{HD 169191}& 4283& 1.88&$-0.03$& 1.4&0.57 &2.35&$-$0.01& 2.68& 1\\
\object{HD 169916}& 4750& 2.50&   0.10& 1.8& 0.69&2.32&$-$0.05& 2.80& 9\\
\object{HD 190421}& 3619& 0.00&$-0.10$& 1.7& 0.58&2.35&$-$0.03& 2.40& 5\\
\object{HD 198357}& 4041& 1.06&$-0.12$& 2.0&0.45 &2.45&  0.01& 2.55& 1\\
\object{HD 199642}& 3912& 0.71&   0.00& 1.6& 0.50&2.43&$-$0.10& 2.50&  1\\
\object{HD 202320}& 4472& 1.74&   0.04& 1.5&0.58 &2.53&$-$0.08& 2.52& 1\\
\object{HD 203638}& 4532& 1.90&   0.15& 1.6&0.69 &2.70&$-$0.07& 2.64& 10\\
\object{HD 210066}& 4118& 1.43&   0.30& 2.0& 0.62&2.75&$-$0.11& 2.90& 1\\
\object{HD 219215}& 3700& 1.00&   0.25& 1.8&0.57 &2.80&$-$0.15& 2.65& 1\\
\object{HD  320868}& 4080& 1.77&   0.05& 1.6&0.59 &2.30&$-$0.08& 2.65& 1\\
\object{2MASS J15023844-4156105}& 4202& 2.13& $-$0.18& 1.6& 0.47&2.35&$-$0.11& 2.56&  1\\

\noalign{\smallskip}
\hline                  
\end{tabular}
\tablefoot{$^{(a)}$ Abundances of Rb and Zr are given on the scale $\log{N({\rm H})} \equiv 12$. $^{(b)}$ Reference for the stellar parameters: (1) AMBRE \citet{wor16}; (2) \citet{hei15}; (3) \citet{kol12}; (4) \citet{luc15}; (5) \citet{mac12b}; (6) \citet{jon17};
(7) \citet{mel08}; (8) \citet{par13}; (9) \citet{alv15}; (10) \citet{kor13}.}
\end{table*}

\begin{acknowledgements}
We acknowledge financial support from the Agencia Estatal de Investigaci\'on of the Spanish Ministerio de Ciencia e Innovaci\'on through the FEDER founds projects 
  PGC2018-095317-B-C21 and PGC2018-102108-B-I00.   
The french coauthors of this article acknowledge financial support from the ANR 14-CE33-014-01 and the "Programme National de Physique Stellaire" (PNPS) of CNRS/INSU co-funded by CEA and CNES. SK acknowledge financial support from the RFBR and Republic of Crimea, project  20-42-910007. We would like to thanks to L. Piersanti and O. Straniero for the discussion on the gravitational settling. Finally, part of the AMBRE parametrisation has been performed with the high-performance computing facility SIGAMM, hosted by OCA.
\end{acknowledgements}

%
%
\bibliographystyle{aa}
\bibliography{rb}

\end{document}